\documentclass[10pt,journal,final,twoside]{IEEEtran}

\IEEEoverridecommandlockouts
\usepackage{mysymbol}

\usepackage{xspace,empheq,fancybox,amsmath,amssymb,graphicx,epstopdf,epsfig,syntonly,times,amsthm} \usepackage{psfrag,color,bm,array,cite}
\usepackage{url,cite,footnote,xspace,syntonly,algorithm,algorithmic}
\usepackage{verbatim,multirow}
\usepackage[T1]{fontenc}
\usepackage{mathtools}
\usepackage{amsmath,amsfonts,amsthm,bm} 
\usepackage{graphicx}
\usepackage{mwe}
\usepackage{caption}
\usepackage{cite}
\usepackage{amsmath,amssymb,amsfonts}
\usepackage{algorithmic}
\usepackage{graphicx}
\usepackage{textcomp}
\usepackage[caption=false,font=footnotesize]{subfig}
\usepackage{color}
\usepackage{xcolor}
\usepackage{cite}
\usepackage{graphicx}

\usepackage{epstopdf}
\usepackage{comment}
\usepackage{multirow}
\usepackage{paralist}
\usepackage{lineno}
\usepackage{mdwlist}
\usepackage{eurosym}\DeclareGraphicsExtensions{.pdf,.png,.jpg}
\usepackage{breqn}
\usepackage{makecell}
\usepackage{soul}
\usepackage{empheq}
\graphicspath{{./pic/}}
\usepackage[super]{nth}
\usepackage{subfig}
\usepackage{bm}
\usepackage[normalem]{ulem}
\usepackage{xcolor}
\def\BibTeX{{\rm B\kern-.05em{\sc i\kern-.025em b}\kern-.08em
    T\kern-.1667em\lower.7ex\hbox{E}\kern-.125emX}}

\usepackage{accents}

\newtheorem{remark}{\bfseries Remark}

\newcommand{\yc}{\color{teal}{}}
\newcommand{\hao}{\color{magenta}{}}

\allowdisplaybreaks

\theoremstyle{remark}

\DeclarePairedDelimiterX\set[1]\lbrace\rbrace{#1}

\begin{document}
\title{\huge LACE-S: Toward Sensitivity-consistent Locational Average Carbon Emissions via Neural Representation}

\author{\IEEEauthorblockN{Young-ho Cho},~\IEEEmembership{Student Member, IEEE}, Min-Seung Ko,~\IEEEmembership{Member, IEEE}, and \IEEEauthorblockN{Hao Zhu},~\IEEEmembership{Senior Member, IEEE}

\thanks{\protect\rule{0pt}{3mm} This work has been supported by NSF Grants 2130706 and 2150571.} \thanks{\protect\rule{0pt}{3mm} The authors are with the Chandra Family Department of Electrical \& Computer Engineering, The University of Texas at Austin, Austin, TX, 78712, USA; Emails: {\{jacobcho, kms4634500, haozhu\}{@}utexas.edu}.}
}
\maketitle

\begin{abstract}
Carbon-aware grid optimization relies on accurate locational emission metrics to effectively guide demand-side decarbonization tasks such as spatial load shifting. However, existing metrics are only valid around limited operating regions and unfortunately cannot generalize the emission patterns beyond these regions. When these metrics are used to signal carbon-sensitive resources, they could paradoxically increase system-wide emissions. 
This work seeks to develop a sensitivity-consistent metric for locational average carbon emissions (LACE-S) using a neural representation approach. To ensure physical validity, the neural model enforces total emission balance through an explicit projection layer while matching marginal emission sensitivities across the entire loading region. Jacobian-based regularization is further introduced to capture the underlying partition of load buses with closely aligned generator responses. Moreover, we present a scalable zonal aggregation strategy, ZACE-S, to reduce the model complexity by mapping nodal inputs to predefined market zones. Numerical tests on the IEEE 30-bus system have verified the performance improvements of LACE-S in matching total emissions and their sensitivities over the non-regularized design. Crucially, while spatial load shifting driven by existing metrics often increases the post-shift emissions, the proposed LACE-S metric has led to a reliable reduction of system-wide emissions, demonstrating its excellent consistency with the global emission patterns.
\end{abstract}

\begin{IEEEkeywords}
Market-based carbon emission metrics, neural representation, sensitivity consistency, Jacobian-based regularization, spatial load shifting.
\end{IEEEkeywords}

\section{Introduction}

Electric power grids are among the largest sources of greenhouse gas emissions, as a critical infrastructure to drive the rapid growth of societal and industrial energy demands~\cite{poudyal2023resilience}. The electricity sector alone produces approximately 40\% of global carbon dioxide (CO$_2$) emissions~\cite{IEA2023}, motivating a transition from purely cost- and reliability-driven grid operations to sustainability-aware paradigms.
Although renewable integration~\cite{pinson2023distributionally,zhao2023uncertainty,ma2025optimizing} and carbon-aware dispatch~\cite{sang2023encoding,chen2024carbon,cho2026carbon} can reduce generation-side emissions, current sustainability goals cannot be met without demand-side adjustments in when and where electricity is consumed. The latter makes it imperative to design locational carbon emission metrics that can quantify the emission responsibility of consumption at each bus~\cite{mollaeivaneghi2025impact}. These metrics, if effectively designed,  allow participants to compare the carbon impact of alternative consumption patterns and encourage shifts toward lower-emission locations, achieving a successful reduction of system-wide emissions.

While generators have well-defined emission factors based on their own fuel types~\cite{cho2026pglib}, this is not the case for load demands. The networked configuration of power grids requires one to allocate system-wide generation emissions to individual loads, for which no unified attribution mechanism currently exists. Existing approaches are broadly categorized into flow-based and market-based metrics. Flow-based metrics, such as the carbon emission flow (CEF) model~\cite{kang2012carbon,kang2015carbon}, treat power flows as carriers of emissions and proportionally distribute nodal emissions across outgoing power lines. However, this proportional sharing principle is mainly an analytical concept and can yield counterintuitive allocations due to complex grid topology such as loop flows~\cite{wang2023carbon}. It also does not incorporate any market-based dispatch information, and thus it can fail to capture marginal sensitivities under binding congestion~\cite{chen2024towards}. As a result, using CEF-based metrics for demand-side adjustment very likely increases the total emissions. Market-based metrics instead compute locational emission factors according to the dispatch models. By mirroring how locational marginal prices (LMPs) incentivize price-responsive demand toward cost-efficient operation~\cite{conejo2005locational}, they can potentially capture the total emissions across different loading conditions.

%
However, existing market-based metrics are insufficient to provide globally consistent emission allocations due to their use of limited information on the operating conditions.
The popular \textit{point-wise} metric, locational marginal carbon emissions (LMCE), analytically computes the marginal sensitivities of total emissions at any operating point~\cite{ruiz2010analysis}. Due to its marginal nature, LMCE cannot be used to allocate carbon emissions or capture the emission pattern beyond the given operating point~\cite{he2021using}.
To tackle this issue, a trajectory-based metric has been developed in~\cite{lu2024market}, namely, the locational average carbon emission using Riemann-sum (LACE-R), by integrating marginal sensitivities along a specific loading path.
But LACE-R is still limited to one loading path and cannot generalize well to a wider range of operating points outside this selected path~\cite{lu2026convex}. For example, the marginal sensitivities provided by LACE-R can be inaccurate for any load shifts not along the loading path.
Thus, a critical gap remains for designing a metric that can capture global emission allocation patterns over the entire loading region while matching the marginal sensitivities for all possible directions of load shifts.

%
%
%

To address these limitations, we develop a sensitivity-consistent metric for locational average carbon emissions (LACE-S) via a neural representation that seeks to generalize over the entire loading region. Specifically, we employ a neural network (NN) to learn a continuous mapping from nodal load profiles to locational emission factors.
By incorporating all possible directions of load shifts, our proposed LACE-S could provide a globally consistent mapping and effectively guide demand-side adjustments that existing metrics fail to support.
To match the nonlinear, complex system-wide emission patterns induced by network-constrained market dispatch, we impose two objective losses on the neural LACE-S modeling. First, an explicit projection layer enforces the total emission balance so that the resultant nodal allocations can be aggregated to match the total emissions. Second, a gradient-based loss aligns the LACE-S model with the marginal emission sensitivities to nodal load perturbations, preserving the first-order consistency with dispatch-induced responses. In addition, we introduce Jacobian-based regularization terms for the neural LACE-S model in order to capture the underlying partition of load buses with closely aligned generator responses.
To further support large-scale implementation, we present a scalable zonal aggregation strategy (ZACE-S) that maps nodal inputs to predefined market zones to reduce model complexity.

The primary value of developing new carbon metrics, as emphasized earlier, lies in supporting demand-side adjustment for carbon-aware grid operations. One particularly important use case is spatial load shifting (SLS)~\cite{lindberg2021guide,lindberg2022using,gorka2025electricityemissions}, by relocating flexible demands across load buses to encourage cleaner generation. Since the load owners do not know the exact market response to possible shift actions, practically implementing SLS requires the use of carbon metrics as signals to estimate carbon reduction potentials. Due to the aforementioned limitations, existing point-wise or trajectory-dependent metrics provide inaccurate carbon estimates and often paradoxically increase the post-shift carbon emissions. The proposed LACE-S representation can overcome this critical gap by providing a globally consistent mapping that more accurately captures the change of carbon emissions over the entire loading region. We validate this capability of LACE-S on the IEEE 30-bus test case~\cite{IEEE_case_ref}, demonstrating that it can reliably guide SLS operations to achieve emission reductions where conventional metrics fail.
In a nutshell, our main contributions are three-fold:
\begin{itemize}
\item We develop a sensitivity-consistent metric, LACE-S, that can explicitly ensure globally consistent allocation by overcoming the limitations of existing metrics. 
\item We put forth a new zonal metric counterpart based on predefined market zones that can achieve scalable computation by reducing the complexity of neural models.
\item Our proposed metric acts as an effective demand-side signal for carbon-aware operations, enabling spatial load shifting that consistently achieves emission reductions across diverse operating conditions.
\end{itemize}

The rest of the paper is organized as follows. Section~\ref{sec:CEM} provides an overview of existing market-based metrics and illustrates their limitations via a motivating example. Section~\ref{sec:LACE} discusses the proposed LACE-S formulation,  along with the scalable counterpart of ZACE-S. Section~\ref{sec:sim} presents numerical validations on the IEEE 30-bus system to demonstrate the modeling accuracy and scalability improvement by ZACE-S. Most importantly, the effectiveness of our proposed LACE-S as SLS signals is verified by consistently achieving emission reductions. Concluding remarks and future research directions are discussed in Section~\ref{sec:con}.

\section{Market-based Carbon Emission Metrics} \label{sec:CEM}
We first present the market-based carbon emission metrics and highlight their limitations, which our proposed model addresses later on.
%
Consider a transmission system consisting of $G$ generators and $D$ load demands, where the generation and load demand are collected in vectors $\bbg\in \mathbb{R}^G$ and $\bbd\in \mathbb{R}^D$, respectively.
The market-clearing process determines the optimal dispatch $\bbg^*$ to serve the given load $\bbd$.
For simplicity, we consider the linear programming (LP) formulation under a given linear cost vector $\bbc$, as follows:
\begin{subequations}\label{opf}
\begin{align}
    \min_{\underline\bbg \leq \bbg \leq \bar\bbg} ~ & \bbc^\top \bbg \\
    \textrm{s.t.} ~
    & \bbA \bbg= \bbb(\bbd),\label{opf_eq}\\
    & \bbl(\bbd) \leq \bbB \bbg \leq \bbu(\bbd).\label{opf_ineq}
 \end{align}
\end{subequations}
The equalities in \eqref{opf_eq} and inequalities in \eqref{opf_ineq} are parameterized by the load $\bbd$ via $\bbb(\bbd)$ and $[\bbl(\bbd),\bbu(\bbd)]$, respectively.
Together, they capture operational constraints such as power balance, line flow limits, and individual unit constraints.
While it is possible to consider quadratic cost functions, nonlinear flow constraints, or flexible loads~\cite{zhou2025machine}, the simple LP formulation is sufficient for capturing the fundamental relationship between $\bbd$ and $\bbg$.

Using the dispatch $\bbg^*$, one can directly compute the system-wide emissions by forming the generator-wise emission vector of length $G$, given by
\begin{align}
    \bbe =\bbf \circ \bbg^*,
    \label{eq:genEmission}
\end{align}
with vector $\bbf\in \mathbb{R}^{G}$ collecting the predefined carbon emission factors of all generators and $\circ$ denoting the entry-wise multiplication. The total system-wide emissions for meeting $\bbd$ become
\begin{align}
    E= \mathbf{1}^\top \bbe = \bbf^\top \bbg^*,
    \label{eq:totEmission}
\end{align}
which helps to define some emission metrics as below. 

\noindent \textbf{ACE: average carbon emissions.} By normalizing $E$ with the total demand based on the load vector $\bbd$, we define a system-wide metric, ACE~\cite{sotos2015amendment}, as given by
\begin{align}\label{eq:ace}
    \eta(\bbd):=\frac{E}{\mathbf{1}^\top \bbd}.
\end{align}
ACE can provide the emission rate per unit load, but it treats all loads as the same and does not assess the locational impact.

\noindent \textbf{LMCE: locational marginal carbon emissions.} This metric is widely used for evaluating the impact from small load changes at a specific bus.
As we know, the market-based response to a small load change is typically limited to a few generators, namely the \textit{marginal generators}, based on network constraints or generator limits. Quantifying this response can effectively provide a location-specific measure of emission responsibility~\cite{valenzuela2023dynamic}.
Let $\mu_i(\bbd)$ denote the bus $i$'s LMCE for a given $\bbd$, as defined by 
\begin{align}
\mu_i(\bbd) := \frac{\partial E}{\partial d_i} = \bbf^\top \cdot \frac{\partial \bbg^*}{\partial d_i}\label{eq:lmce}
\end{align}
where $\frac{\partial \bbg^*}{\partial d_i}$ can be computed analytically via the Lagrangian sensitivity analysis of~\eqref{opf}; see e.g.,~\cite{lu2024market}.
This sensitivity-based metric also has some limitations.
First, the LMCE values can change sharply with varying operating points, as the marginal generators depend on the market congestion condition. More critically, LMCE only quantifies the \textit{marginal impact} which makes it unable to assess the \textit{aggregated emission effects} of individual loads.

\noindent \textbf{LACE: locational average carbon emissions.} This metric has been introduced with the goal of capturing the average, or \textit{aggregated impact} of an individual load. To this end, one recent attempt~\cite{lu2024market} proposes to average the accumulated marginal emissions along a normalized loading path.
This is equivalent to a \textit{Riemann-sum integral} by summing up the impact from each small step along the path. Thus, it will be termed LACE-R, as given by
\begin{align}
\lambda_i(\bbd) := \frac{1}{d_i}
{\int^1_0 \mu_i(\rho \bbd) d\rho} \label{lace-R}
\end{align}
where the variable $\rho\in[0,1]$ is used to scale $\bbd$ to produce the normalized loading path.
By smoothing $\mu_i$ along the fixed path, LACE-R no longer has the sharp changes like LMCE. More importantly, it can achieve the total emission balance, ensuring that the emissions allocated to individual loads sum up to the total $E$.
However, a fundamental issue still exists for LACE-R due to its dependence on the chosen path in \eqref{lace-R}.
This path dependence significantly limits the generalizability of LACE-R, since it is unable to capture the global emission patterns over the entire loading region beyond the path.
We present a simple example to illustrate these issues.


\begin{figure}[t!]
	\centering
	\subfloat[]{\includegraphics[scale=0.25]{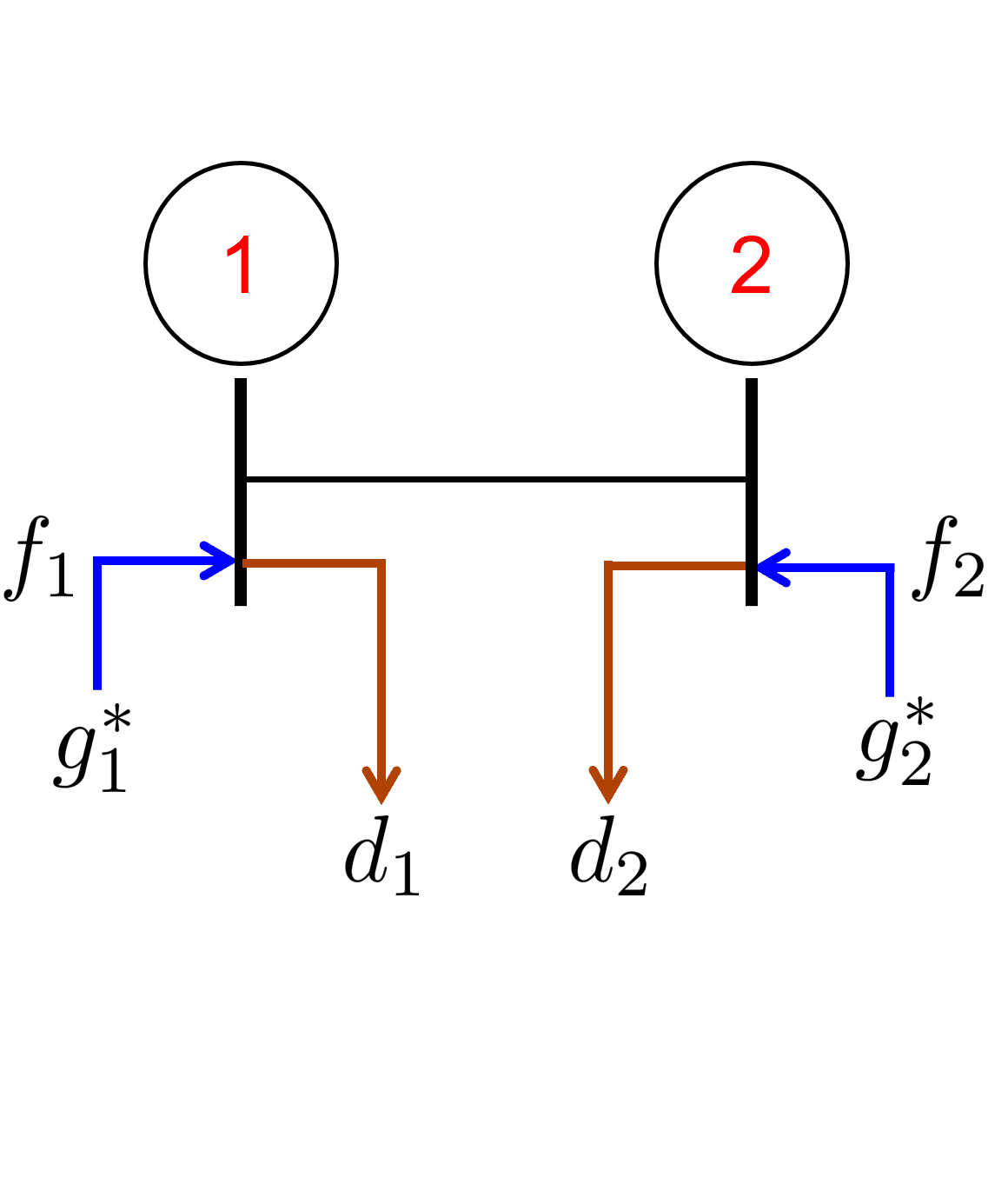}}
    \subfloat[]{\includegraphics[scale=0.25]{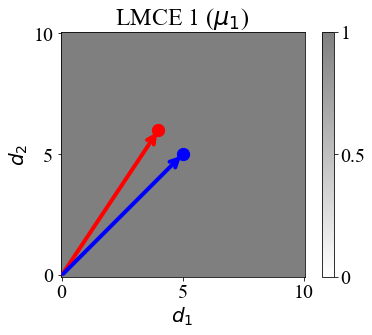}}
    \subfloat[]{\includegraphics[scale=0.25]{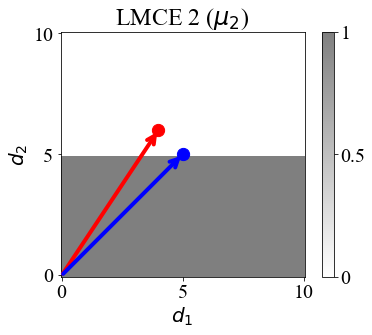}}
	\caption[]{\small (a) A 2-bus system with two generators and loads; and the LMCE distribution of (b) $d_1$ and (c) $d_2$.}\label{example}
\end{figure}

\subsection{Limitations of Existing Metrics} \label{sec:CEM_mot}
 
We will show the limitations of existing metrics in terms of balancing the total emissions and matching the relative emission responsibility among individual loads. 
Consider a simple two-bus system depicted in Fig.~\ref{example}(a) with a connecting line capacity of 5 MW. Assume $G_1$ uses carbon-intensive fuel ($f_1=1$ tCO$_2$/MWh) with an incremental cost of $c_1$ while $G_2$ is carbon-free and powered with clean resources ($f_2=0$ tCO$_2$/MWh) at a higher cost of $c_2>c_1$.
Although carbon-free resources tend to be cheaper than fossil-fueled ones in practice, we adopt a cheap–dirty versus expensive–clean setting which is helpful for creating a complex dispatch-emission relation in the two-bus system.  
Figs.~\ref{example}(b) and (c) respectively illustrate the LMCE values of $\mu_1$ and $\mu_2$ across the loading region of $(d_1,d_2)$. Any change in $d_1$ can be met by the cheaper $G_1$, but once $d_2$ exceeds the 5-MW line capacity,  its change has to be met by the more costly $G_2$. The latter leads to a distinct value of $\mu_2=0$, while the LMCE value everywhere else is 1. This LMCE pattern is better aligned with that in large, complex, real-world power grids, where reducing emissions could incur additional generation costs.

We consider two operating points: $(d_1,d_2) = (5,5)$ MW with the economic dispatch of $(g_1,g_2)=(10,0)$ MW and $(d_1,d_2) = (4,6)$ MW with $(g_1,g_2) = (9,1)$ MW due to line congestion. Hence, shifting 1 MW of load from $d_1$ to $d_2$ can reduce the total emissions $E$ from 10 to 9 tCO$_2$/h. If a locational  metric would assign a higher emission factor to $d_1$ than $d_2$, it can better match the difference in emission responsibility between the two loads and thus shift the loads toward more carbon reduction.

We first show the limitations of LMCE. For the $(4,6)$ MW loading, its LMCE values of $(\mu_1,\mu_2)=(1,0)$ give rise to
\begin{align*}
\mu_1 \cdot d_1 + \mu_2 \cdot d_2 = 4 < E=9, 
\end{align*}
which significantly underestimates the total emissions. 
As for the $(5,5)$ MW loading, the equal LMCE values of $\mu_1 = \mu_2=1$ incidentally match the total $E=10$, but fail to recognize the difference between the two loads in reducing emissions. Fundamentally, LMCE is a marginal metric obtained by local information only, making it ineffective for quantifying the aggregated impact of individual loads or signaling the carbon reduction potentials.  

The issue of limited information remains for the LACE-R metric, although it is able to achieve total emission balance. Using the paths shown in Fig.~\ref{example}, the integral in \eqref{lace-R}  produces LACE-R values of $(\lambda_1,\lambda_2) = (1, 0.833)$ at the $(4,6)$ MW loading. Thus, $\lambda_1\cdot d_1 + \lambda_2 \cdot d_2 = 9$, correctly matching the total $E=9$.
But still,  the $(5,5)$ MW loading has equal values of $(\lambda_1,\lambda_2) = (1, 1)$, which are exactly the same as LMCE. Although LACE-R is consistent for carbon accounting by accumulating the information along the loading path, it is still ineffective for exploring carbon reduction potentials outside the path. 
As demonstrated by our numerical tests in Section \ref{sec:sls}, LACE-R tends to produce load shifts similar to those of LMCE, suggesting that the former is limited to the integration path only and unable to capture the broader emission patterns.

These observations have confirmed the aforementioned limitations of existing market-based emission metrics. Interestingly, our proposed LACE metric (introduced in the ensuing section) yields $(\lambda_1,\lambda_2) = (1.013, 0.987)$ at $(d_1,d_2) = (5,5)$ MW, and $(\lambda_1,\lambda_2)=(0.967, 0.855)$ at $(4,6)$ MW. They not only achieve the total emission balance, but also match the difference between the two loads in carbon reduction at $(5,5)$ MW. With  $\lambda_1 > \lambda_2$, our proposed metric encourages the load shift from $d_1$ to $d_2$ for reducing emissions, making it effective for signaling the carbon reduction potentials.

\section{\textbf{LACE-S} via Neural Representation} \label{sec:LACE}

We propose a new LACE metric that seeks to achieve consistency with the emission patterns across the entire loading region. We term it \textit{\textbf{LACE-S}}, using a neural network (NN)-based representation for the LACE function $\bblambda(\bbd)$ to fit the sensitivity of emissions everywhere. 
As discussed earlier, the fundamental issue of LACE-R stems from the limited path-only information that is available to obtain this metric. Interestingly, LACE-R could be viewed as a NN formed by data points solely sampled from the \textit{single loading path}, as NNs can accurately represent continuous functions such as integrals~\cite{hornik1989multilayer}.  
Inspired by this viewpoint, we put forth the NN-based LACE-S with training data points spanning the entire loading region, making it more effective in capturing the emission patterns throughout the loading region.

\begin{figure}[t!]
	\centering
	\includegraphics[scale=0.35]{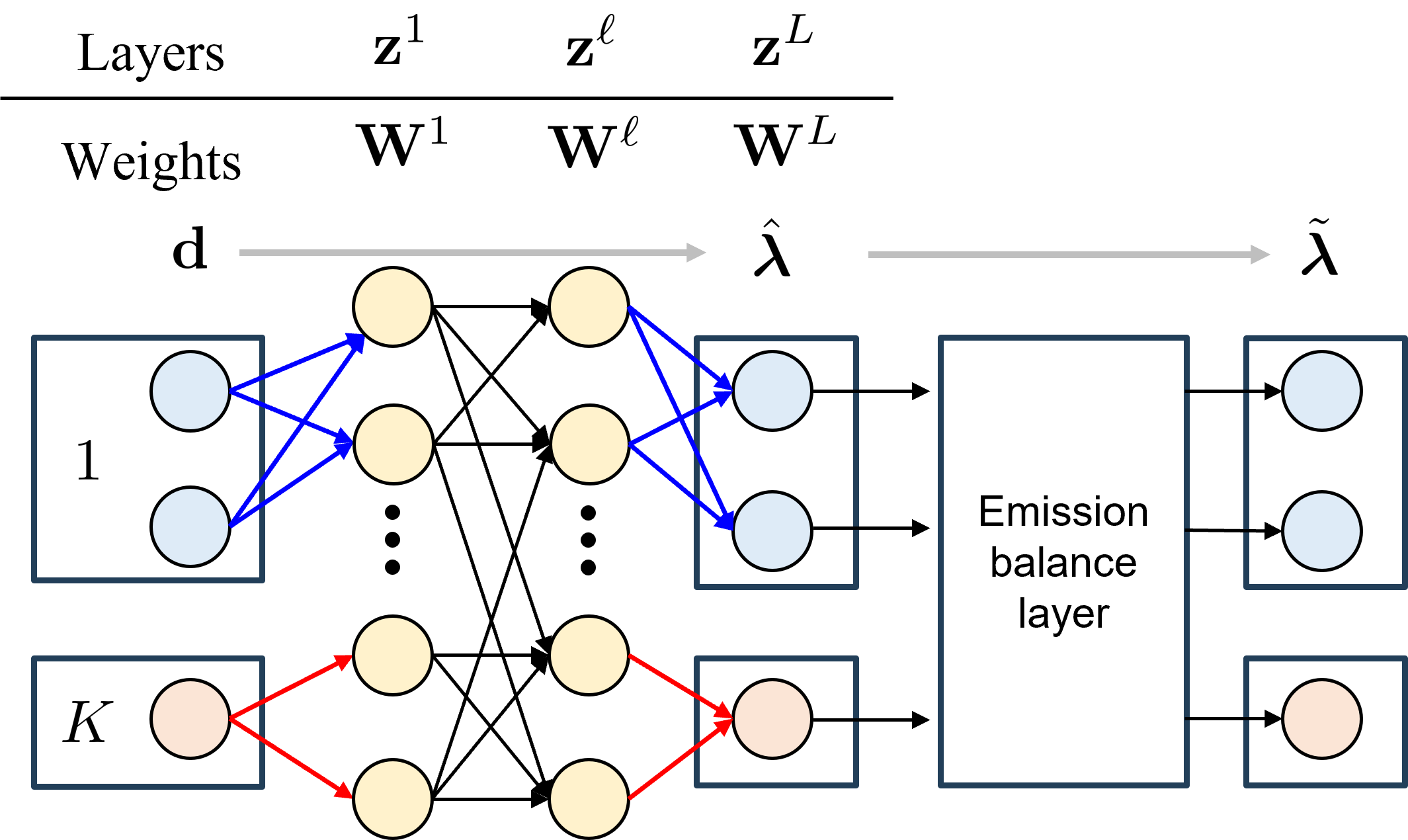}
	\caption[]{\small The proposed neural approximation models estimate LACE-S from load profiles using trainable weights. To mitigate overfitting, cluster-based structured sparsification is applied to the first and last layers.}\label{NN_LACE}
\end{figure}

Fig.~\ref{NN_LACE} illustrates the overall architecture of our proposed feedforward NN of $L$ hidden layers to represent the LACE vector $\hat \bblambda \in \mathbb R^D$ for any load profile $\bbd \in \mathbb R^D$.
Denoting the layer $\ell$ by $\bbz^{\ell}$, with the input $\bbz^{0}=\bbd$ and the output $\bbz^{L}=\hat \bblambda$, we form the layer-wise transformation as
\begin{align}
    \bbz^{\ell} &= \sigma^{\ell}(\bbW^{\ell} \bbz^{\ell-1} + \bbb^{\ell}),\quad \ell=1,\dots,L, \label{NN}
\end{align}
where $\sigma^{\ell}(\cdot)$ stands for the activation function with corresponding trainable weights and biases in $\{\bbW^{\ell},\bbb^{\ell}\}$.
We apply ReLU activation for the first $(L-1)$ layers and the Sigmoid function for the output layer $L$.  The latter has been chosen to ensure that each component of $\hat \bblambda$ stays within the normalized range of $[0, 1]$. Since LACE values are inherently bounded by the minimum and maximum values of $\bbf$, or the range of system-wide generator emission factors, this Sigmoid activation is useful for bounding the output  $\hat \bblambda$.

We also configure the NN connectivity to reflect the modularity in typical power networks. In general, loads with similar LMCE patterns are served by the same set of marginal generators, and thus their LACE values are more strongly coupled. Inspired by this, we can partition the buses into clusters with similar LMCE values and apply this partition to sparsify the first and last layers, namely $\bbW^1$ and $\bbW^L$. This block-sparse layer design keeps any entries that link different clusters to be zero. For the other layers, we also apply dropout to reduce the NN model complexity. Our sparsified NN architectural design not only can mitigate data overfitting by leveraging the modularity of power systems, but also improves the computation efficiency and numerical stability during NN training; as corroborated by earlier work on NN-based grid modeling~\cite{cho2024data,cho2025sparse}.
The modularity of power systems will be further explored to design zone-based emission metrics via aggregating multiple loads into a zone, as detailed in Section~\ref{sec:ZACE}.

To serve as a valid LACE metric, the neural-represented values must strictly ensure they correctly disaggregate the total emissions. Recall that $E = \bbf^\top \bbg^*$ in \eqref{eq:totEmission} where the dispatch $\bbg^*$ depends on the given load $\bbd$. While enforcing the total emission balance $\bbd^\top \tdbblambda=E$ is essential, it is difficult to achieve by relying solely on the hidden layers. To address this, we add an emission balance layer to convert $\hhatbblambda$ to the final LACE output $\tdbblambda$ that matches the true total.
This layer implements a projection onto the hyperplane $\mathcal{H}$, as given by
\begin{align}
    \mathcal{H} = \{ \bblambda \in \mathbb{R}^D ~|~\bbd^\top \bblambda = E\}.
\end{align}
Solving this projection yields a linear transformation, as 
\begin{align}
   \tilde \bblambda = \hat \bblambda - \frac{\bbd^\top \hat \bblambda - E}{ \| \bbd \|^2_2 } \bbd = \left(
   \bbI - \frac{\bbd\bbd^\top}{\| \bbd \|^2_2 } \right) \hat{\bblambda} + \frac{ E}{ \| \bbd \|^2_2 } \bbd, \label{projection}
\end{align}
which is both computationally and analytically tractable. 
In addition, this projection shows that the difference term $(\hat \bblambda - \tilde \bblambda)$  can be used to quantify the deviation from the perfect balance condition, motivating us to define the \textit{balance loss} as
\begin{align}\label{eq:balanceloss}
    \mathcal{L}_{\lambda} = \| \hat \bblambda - \tilde \bblambda \|^2_2 = \left\| \frac{\bbd^\top \hat \bblambda - E}{ \| \bbd \|^2_2 } \bbd \right\|_2^2 = \frac{(\bbd^\top \hat \bblambda - E)^2}{\|\bbd \|^2_2}.
\end{align}
Minimizing $\ccalL_\lambda$ encourages the NN output $\hhatbblambda$ to closely approach the emission balance condition and thus reduce the level of distortion due to projection.

In addition to the emission balance, another crucial property for LACE outputs is to ensure they capture the LMCE-based sensitivity.
Note that the mapping to $\hhatbblambda$ cannot be trained as a supervised learning task, because the ground-truth LACE values are not known or available as target labels.
Instead, we leverage the known LMCE values as the \textit{first-order sensitivity}, which provides the \textit{gradient label} for training $\hhatbblambda$.
Specifically, the NN for $\hhatbblambda$ provides the estimated LMCE $\hat \bbmu$ per its definition as the partial derivatives of the estimated total emissions, namely $\bbd^\top\hat\bblambda$ [cf.~\eqref{eq:lmce}]:
\begin{align}
    \hat \bbmu = \frac{\partial (\bbd^{\top} \hat \bblambda)}{\partial \bbd}.
\end{align}
Note that we have used $\hhatbblambda$ to derive $\hhatbbmu$ instead of its projection $\tdbblambda$, because the latter satisfies the balance condition and will not produce the LMCE estimates by the NN.
Using the estimated $\hhatbbmu$, we define the \textit{sensitivity loss} as 
\begin{align}
    \mathcal{L}_{\mu} = \| \hat \bbmu - \bbmu \|^2_2.
\end{align}
The objective for training  $\hhatbblambda$ combines both the balance loss $\mathcal{L}_{\lambda}$ and sensitivity loss $\mathcal{L}_{\mu}$ into:
\begin{align}
    \mathcal{L} = \mathcal{L}_{\lambda} + \mathcal{L}_{\mu}.\label{loss_L}
\end{align}
This dual objective can ensure sensitivity consistency with known marginal emissions everywhere while respecting the system-wide emission balance.

\begin{figure}[t!]
	\centering
	\subfloat[Congestion 1]{\includegraphics[scale=0.22]{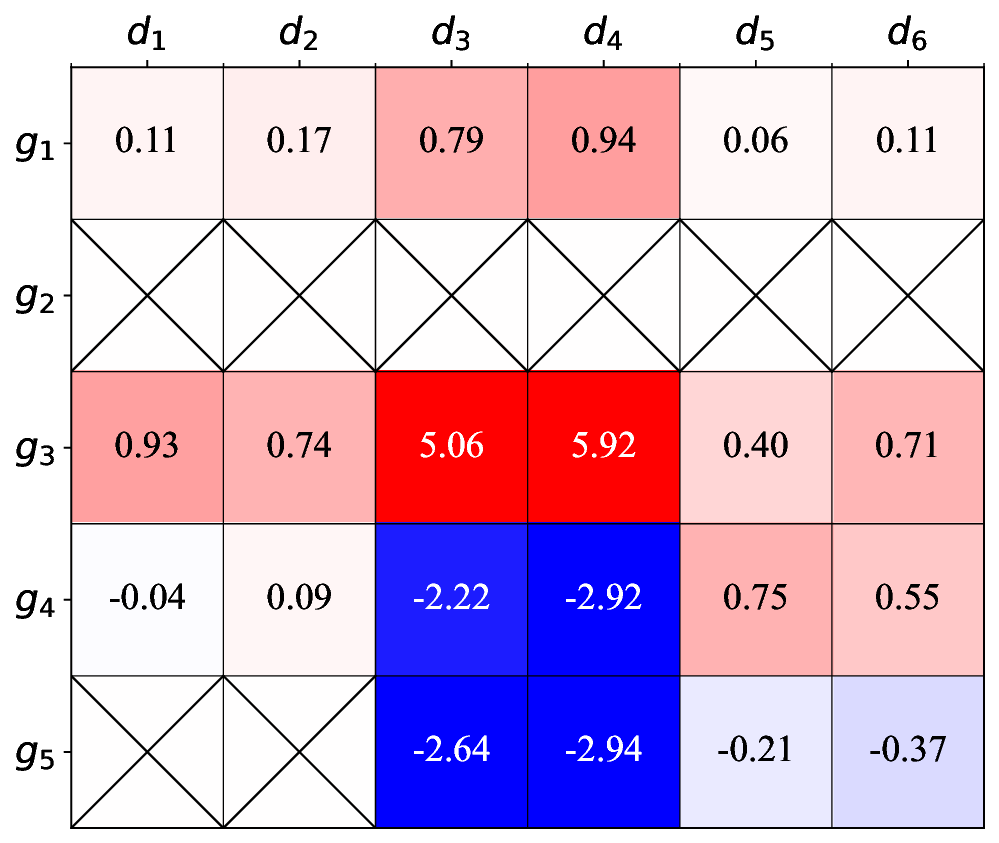}}\quad
    \subfloat[Congestion 2]{\includegraphics[scale=0.22]{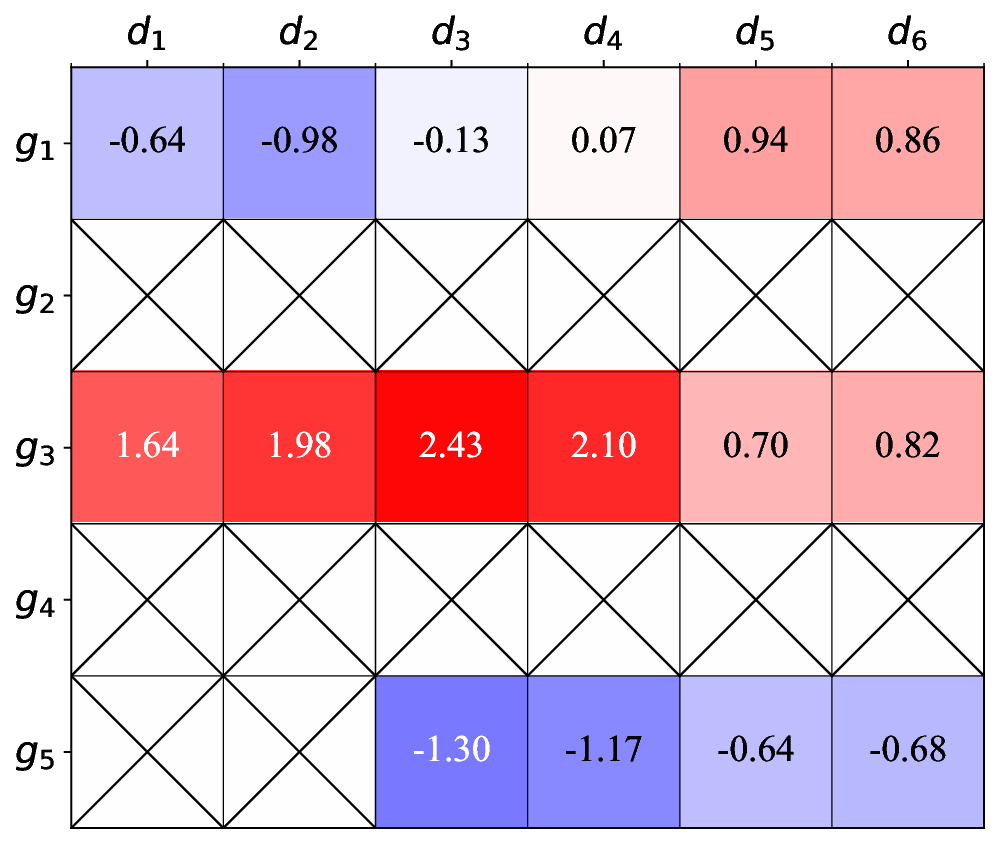}}
	\caption[]{\small Visualization of dispatch Jacobian matrices ($\partial \bbg^*/\partial \bbd$) on the IEEE 14-bus system under two distinct congestion patterns.}\label{fig:congest_pattern}
\end{figure}

\subsection{Jacobian-based Regularization} \label{sec:LACEreg}
To ensure that the learned mapping respects the underlying market structure, we introduce regularization terms on the Jacobian matrix of our LACE-S model, as defined by $\bbJ := [\partial \hat{\lambda}_i (\bbd) / \partial d_j]$.
In market dispatch, load buses can be partitioned into groups that consistently share similar marginal generator responses across varying operating points~\cite{valenzuela2023dynamic,ji2016probabilistic}.
Specifically, a small load perturbation at any bus mainly affects only emission factors of those buses within the same group, with minimal effects on buses outside the group.
Although this partition of load buses is consistent with the aforementioned block-sparse NN layer design, the latter alone is insufficient to promote this partitioned structure for the overall NN mapping produced by multiple layers.
Hence, we will directly regularize the LACE-S model's Jacobian matrix $\bbJ$, which exactly quantifies the marginal effect of location-specific load perturbations.

%
%

\begin{remark}{(Example of Load Bus Partition.)} \label{rmk:congest}
Empirical analysis of the dispatch Jacobian matrix $\frac{\partial \bbg^*}{\partial \bbd}$ confirms this partitioning behavior among load buses. We evaluated the IEEE 14-bus system with tightened line limits to induce diverse congestion patterns~\cite{kekatos2014grid}, applying linear cost functions to maintain a constant Jacobian matrix within each pattern.
As illustrated in Fig.~\ref{fig:congest_pattern}, the generator responses naturally form three distinct clusters $\{d_1, d_2\}$, $\{d_3, d_4\}$, and $\{d_5, d_6\}$.
Within each cluster, marginal responses are closely aligned, whereas the distinct boundaries between blocks confirm the weak inter-cluster coupling. This intra-cluster similarity remains consistent even when the congestion pattern changes. This partition of load buses generally exists in practical systems \cite{valenzuela2023dynamic,ji2016probabilistic}, which will be utilized by our proposed regularization design to promote the structural coupling.
\end{remark}

To this end, we introduce two regularization terms for the Jacobian matrix $\bbJ$  to encourage block-diagonal and diagonally dominant structures. The block-diagonal structure follows from the congestion-based clusters and can be promoted using an $\ell_1$-norm term:
\begin{align}
    \mathcal{L}_{\mathrm{bd}}=\|\bbJ-\operatorname{bdiag}(\bbJ)\|_1,\label{loss_gamma1}
\end{align}
where $\operatorname{bdiag}(\cdot)$ retains entries ($i,j$) only when loads $i$ and $j$ belong to the same cluster.
As for the diagonal dominance, it seeks to further sparsify the model  to encourage the self-influence term ($\partial \hat \lambda_i/\partial d_i$) to be much larger than other cross-influence terms ($\partial \hat\lambda_i/\partial d_j$ for $j\neq i$). We achieve this using a relaxed $\ell_1$ term:
\begin{align}
    \mathcal{L}_{\mathrm{d}}= \Big \| \Big (|\bbJ-\operatorname{diag}(\bbJ)| - \varepsilon \Big )_{+} \Big \|_1 \label{loss_gamma2}
\end{align}
where $\operatorname{diag}(\cdot)$ retains all diagonal entries with $(\cdot)_{+}$ projecting to the positive orthant like the ReLU. The tolerance parameter $\varepsilon>0$ is pre-determined to match the expected level of intra-cluster couplings: if $|J_{ij}|\leq \varepsilon$, the penalty becomes zero. The overall training objective augments the data-fitting loss $\mathcal{L}$ in \eqref{loss_L} with these two Jacobian-based regularizers to form
\begin{align}
    \mathcal{L}_{\mathrm{reg}} = \mathcal{L} + \gamma_1 \mathcal{L}_{\mathrm{bd}} + \gamma_2 \mathcal{L}_{\mathrm{d}},\label{loss_total}
\end{align}
where hyperparameters $\gamma_1$ and $\gamma_2$ are selected to balance between the modeling accuracy and the structural consistency, as discussed in Remark~\ref{rmk:LACE_reg}.

\begin{figure}[t!]
    \vspace{-0.5em}
	\centering
	\subfloat[]{\includegraphics[scale=0.22]{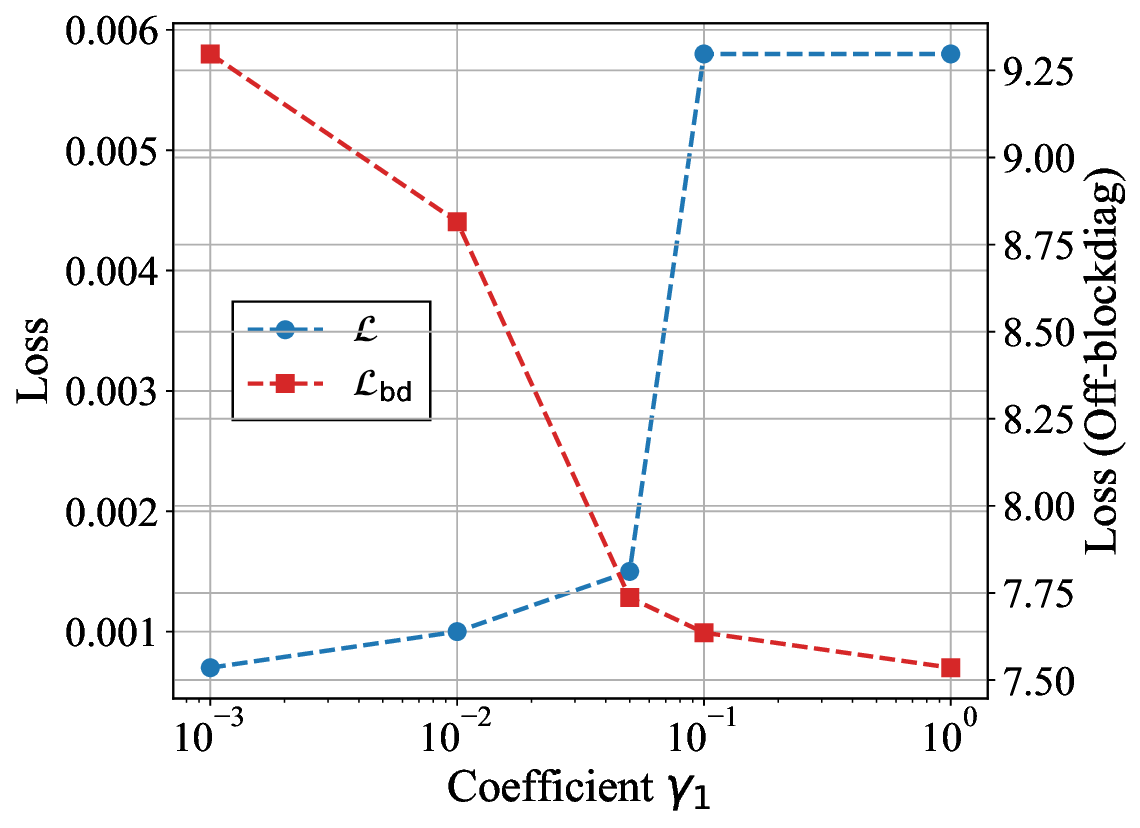}\label{Selection_gamma1}}
    \subfloat[]{\includegraphics[scale=0.22]{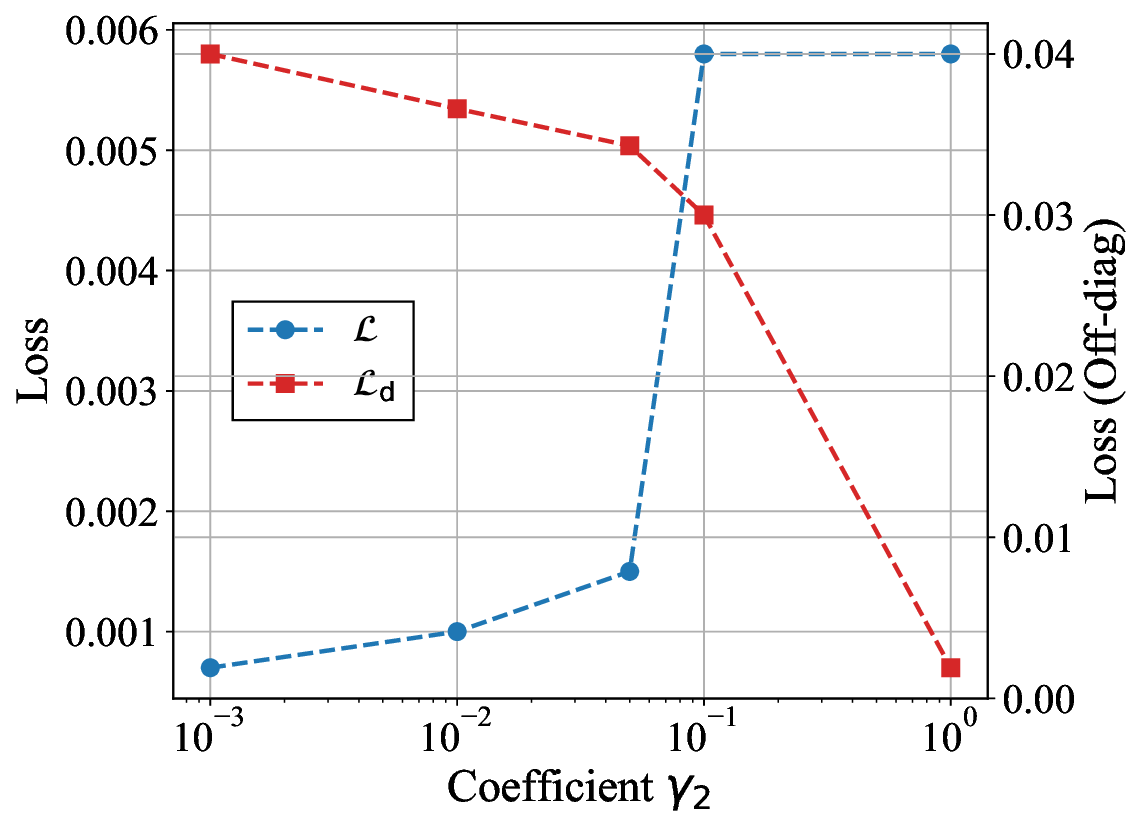}\label{Selection_gamma2}}
	\caption[]{\small Comparison of $\mathcal{L}$ and the regularization losses $\{\mathcal{L}_{\mathrm{bd}},\mathcal{L}_{\mathrm{d}}\}$ on the IEEE 30-bus system. (a) Vary $\gamma_1$ with $\gamma_2$ fixed at 0; (b) Vary $\gamma_2$ with $\gamma_1$ fixed at 0.05.}\label{Selection_gamma}
\end{figure}

\begin{remark}{(Selecting the regularization hyperparameters.)} \label{rmk:LACE_reg}
We determine the regularization coefficients $\gamma_1$ and $\gamma_2$ via a sequential parameter sweep, aiming to identify the best trade-off between approximation accuracy (primary loss $\mathcal{L}$) and structural consistency (regularization terms). By performing a line search along each parameter dimension, we seek a threshold where the Jacobian structure is effectively enforced without causing a large increase in the primary loss.
For example, Fig.~\ref{Selection_gamma} plots the loss curves for the candidate pairs ($\gamma_1, \gamma_2$) on the IEEE 30-bus system. 
We first fix $\gamma_2=0$ and sweep $\gamma_1$, and  Fig.~\ref{Selection_gamma}(a) shows that $\gamma_1=0.05$ offers a sufficient reduction of the block-diagonal loss without compromising the primary loss. Upon fixing $\gamma_1=0.05$, we sweep $\gamma_2$ and Fig.~\ref{Selection_gamma}(b) shows that  $\gamma_2=0.05$ is a good choice. These procedures can be extended to larger systems to ensure the resultant LACE-S model consistently yields high accuracy while enforcing structurally consistent emission attribution.
\end{remark}

\subsection{Training Procedure} \label{sec:procedure}

To optimize the Jacobian-regularized loss $\mathcal{L}_{\mathrm{reg}}$  in \eqref{loss_total}, 
directly minimizing it could be computationally expensive or even numerically unstable. Since the Jacobian matrix includes the NN model's second-order derivatives, optimizing it significantly increases the computation of backpropagation at reduced numerical accuracy. 
To enhance the convergence, we adopt a multi-stage training procedure, in which loss terms are gradually incorporated to promote stable yet fast convergence.
\begin{itemize}
\item \textbf{Stage 1:} Initialize the trainable parameters to predict the known ACE metric in \eqref{eq:ace}. Anchoring the NN output to this baseline provides a stable initialization for the subsequent training stages.
\item \textbf{Stage 2:} Train with the primary loss $\mathcal{L}$ in~\eqref{loss_L} to approach the conditions of total balance and sensitivity consistency. 
\item \textbf{Stage 3:} Add the block-diagonal regularization $\mathcal{L}_{\mathrm{bd}}$ to reduce the cross-cluster couplings.
\item \textbf{Stage 4:} Add the diagonal-dominance penalty $\mathcal{L}_{\mathrm{d}}$ to improve the overall conditioning of the Jacobian $\bbJ$.
\end{itemize}
The training can proceed to the next stage once the current stage has converged, using a per-epoch loss-reduction threshold such as $1$e$-3$.

\subsection{Zonal Average Carbon Emissions (ZACE) for Scalability} \label{sec:ZACE}

For large-scale systems, predicting the full vector $\bblambda$ can be computationally prohibitive, as the number of NN parameters can scale quadratically, $\mathcal{O}(D^2)$, for a system of $D$ load buses. 
Inspired by the popular practice used in most electricity markets, we introduce the zonal average carbon emissions (ZACE) metric 
with $K$ predefined market zones ($K \ll D$) to reduce the parameter complexity to $\mathcal{O}(D \cdot K)$. Hence, this metric
significantly lowers memory requirements for optimizing the Jacobian while producing metrics compatible with zone-based pricing and market operations.

\begin{figure}[t!]
	\centering
    \includegraphics[scale=0.35]{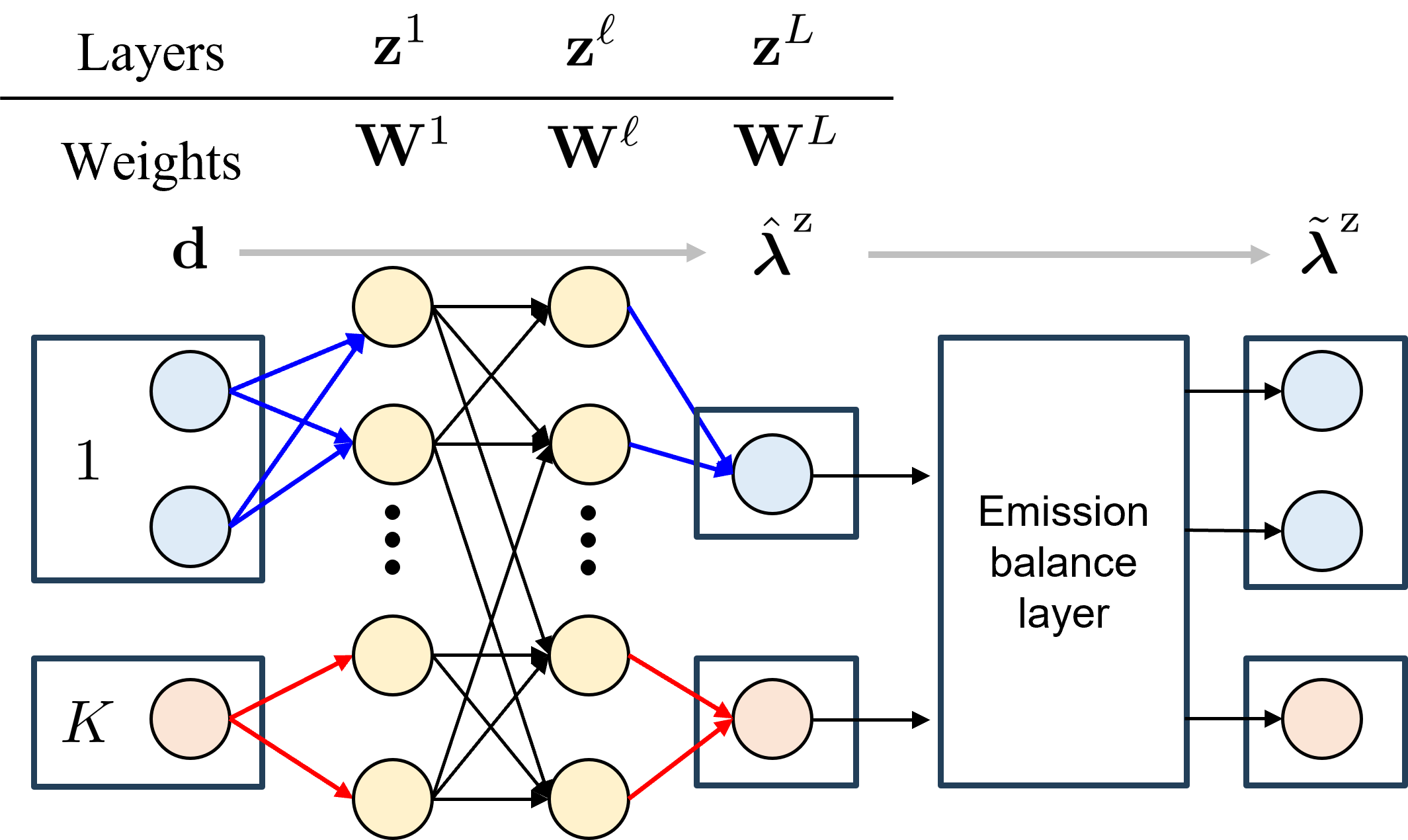}
	\caption[]{\small The proposed neural approximation models estimate ZACE-S from load profiles using trainable weights.}\label{NN_ZACE}
\end{figure}

For the NN to predict the ZACE-S vector denoted by $\hat \bblambda^{\mathrm z}$, it suffices to modify the final layer to match this reduced dimension of $K$. The loss design follows similarly to match the balance and sensitivity of emissions at the zonal level. Specifically, the balance loss is formulated similar to \eqref{eq:balanceloss} by replacing $\bbd$ with the zonal load vector. 
Moreover, we define the zonal marginal carbon emission (ZMCE) using a load-weighted average of the nodal LMCEs within each zone $k$, as 
\begin{align}
\mu^{\mathrm z}_k = \sum_{i \in \mathcal{Z}_k} \left( \frac{d_i}{\sum_{j \in \mathcal{Z}_k} d_j} \right) \mu_i,~\forall k = 1,\ldots, K \label{eq:zmce}
\end{align}
where $\mathcal{Z}_k$ is the set of loads within zone $k$.
This aggregation method follows the zonal marginal price (ZMP) used in the major electricity markets~\cite{ISONEManual11, PJMManual28}. 
This way, the primary loss function $\mathcal{L}^{\mathrm z}$ combines the two losses to match, respectively, the balance and ZMCE vector $\bbmu^{\mathrm z}$. Similar to Section \ref{sec:LACEreg}, it is further regularized by a zone-based sparse structure in the Jacobian matrix $\bbJ^{\mathrm z}\in \mathbb{R}^{K\times D}$,
%
to ensure that each zonal emission factor is mostly sensitive to local load perturbations within its own zone. Specifically, let the binary matrix $\bbM \in \{0,1\}^{K\times D}$ indicate the zone-node connectivity, where $M_{k i} = 1$ if load $i$ belongs to zone $k$ and 0 otherwise. The regularized training objective now becomes:
\begin{align}
    \mathcal{L}^{\mathrm{z}}_{\mathrm{reg}}
    = \| \hat \bblambda^{\mathrm z} - \tilde \bblambda^{\mathrm z} \|^2_2
     + \| \hat \bbmu^{\mathrm z} - \bbmu^{\mathrm z} \|^2_2 
      + \gamma_3 \big\| \bbJ^{\mathrm z}  - \bbM \odot \bbJ^{\mathrm z} \big\|_1, \label{loss_total2}
\end{align}
where $\odot$ denotes the element-wise product to produce the connectivity-based blocks [cf.~\eqref{loss_gamma1}]. The training procedure for the ZACE loss in \eqref{loss_total2} follows similar steps in Section \ref{sec:procedure}. 
While this zonal aggregation provides a coarser representation of emission allocations, it maintains high accuracy at substantially reduced memory requirements, as demonstrated by our numerical results later.

\section{Numerical Validations for Neural Models} \label{sec:sim}
%

We validate the proposed neural LACE-S and ZACE-S models on the IEEE 30-bus test case~\cite{IEEE_case_ref}, which contains 20 loads and 6 generators.
To promote the spatial variation in locational emission signals, we reassign generator fuel types in the original case. Specifically, we map each generator to one of three fuel categories: high-emission anthracite coal (ANT), mid-emission distillate fuel oil (PEL), or low-emission gas combined cycle (CCGT), as detailed in Table \ref{tab:fuel_sim}. Emission factors are quantified using CO$_2$-equivalent (CO$_2$e) intensities, which account for the global warming potential of methane (CH$_4$) and nitrous oxide (N$_2$O) in addition to CO$_2$; see more discussions in \cite{cho2026pglib}. We will first show the excellent performance of our proposed LACE-S metric in matching system-wide emissions and marginal sensitivity, and further demonstrate its applicability to an important carbon-aware grid operational use case, spatial load shifting (SLS), as an effective signal for representing the carbon reduction potentials.

\begin{table}[t!]
\centering
\caption{Generator fuel type assignments}
\label{tab:fuel_sim}
\begin{tabular}{l|cc}
\hline
Fuel Type & CO$_2$e Factors & Bus Index \\
\hline
Gas combined cycle (\textbf{CCGT})        &  0.3625 & \{1\}   \\ 
Distillate fuel oil (\textbf{PEL})        &  0.7018 & \{2, 13\}   \\ 
Anthracite coal (\textbf{ANT})            &  0.9143 & \{22, 23, 27\} \\
\hline
\end{tabular}
\end{table}

To train the proposed NN models, we generate 50,000 samples using a standard DC-optimal power flow (OPF) formulation as the market-clearing problem. 
Note that each sample draws a feasible load scenario $\bbd \in \mathbb R^{20}$ by uniformly scaling the nominal loads within the range of $[110,~130]\%$. For this test case, loading levels below 110\% do not induce market congestion, leading to the same LMCE values everywhere. Hence, this increase of loading is intended to produce non-trivial and spatially differentiated carbon signals. The DC-OPF problem is solved by the MATPOWER simulator~\cite{zimmerman2010matpower} to obtain the optimal generator dispatch $\bbg^*$. Using the latter, we compute the two targets for training the NNs: the total emissions $E$ in \eqref{eq:totEmission} and the LMCE vector $\bbmu$ in \eqref{eq:lmce}. 
Using these samples, the NN modeling has been performed in PyTorch with Adam optimizer on a standard laptop equipped with Intel\textsuperscript{\textregistered} CPU @ 2.70 GHz, 32 GB RAM, and NVIDIA\textsuperscript{\textregistered} RTX 3070 Ti GPU @ 8GB VRAM.
We optimize NN parameters via backpropagation using the loss functions in \eqref{loss_total} and \eqref{loss_total2} for LACE-S and ZACE-S, respectively, for 1,000 epochs with a learning rate $1$e$-3$. We use 90\% of the samples for training and the remaining 10\% for testing, and all reported performance comparisons are evaluated in the testing set.

\subsection{LACE-S Modeling Performance}
We evaluate the proposed LACE-S model along two complementary dimensions: its ability to satisfy the total emission balance and its accuracy in approximating LMCE-based marginal sensitivities.
We compare our proposed LACE-S model incorporating all sparsification and regularization techniques against a fully-connected baseline model denoted by Full\_NN. Both models share the same architecture with three hidden layers of 40 neurons each. Our LACE function mapping $\bbd \rightarrow \bblambda$ has $20$ variables in both inputs and outputs. 
To sparsify the fully-connected weights, we partition the 30-bus system into 4 areas using k-means clustering based on LMCE similarity. The hidden neurons in the first and last layers are then proportionally distributed among these 4 areas. For the regularization coefficients in \eqref{loss_total}, we set $\gamma_1=0.1$ and $\gamma_2=0.01$ following the selection rationale in Remark~\ref{rmk:LACE_reg}.

\begin{figure}[t!]
    \vspace{-0.5em}
	\centering
	\subfloat[Max deviation]{\includegraphics[scale=0.39]{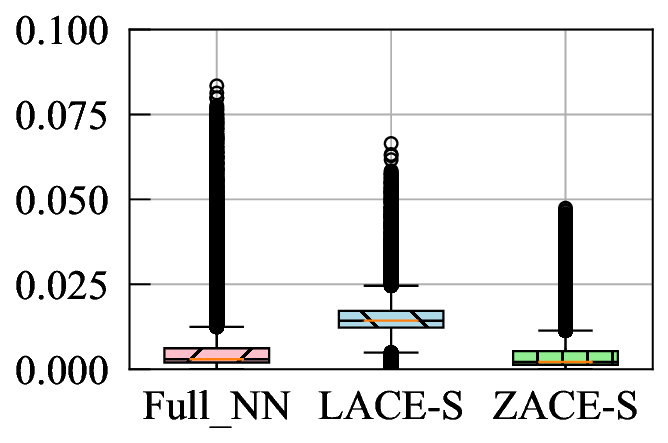}}
    \subfloat[Average deviation]{\includegraphics[scale=0.39]{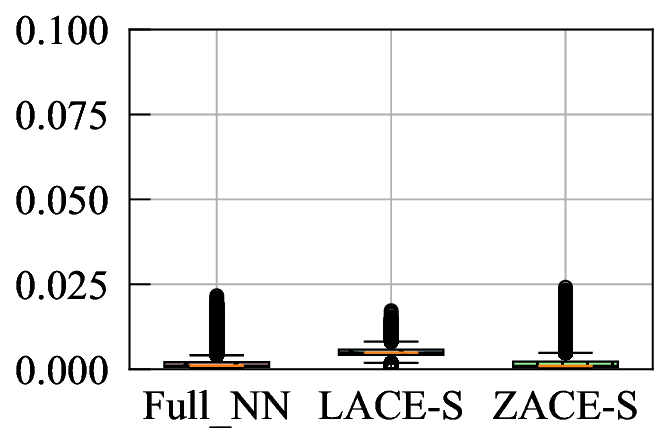}}
	\caption[]{\small Comparisons of the (a) maximum and (b) average absolute deviation from the perfect balance condition for the Full\_NN, LACE-S, and ZACE-S models using the 30-bus system.}\label{LACE_sim_1}
\end{figure}

\begin{figure}[t!]
	\centering
	\subfloat[Max deviation]{\includegraphics[scale=0.39]{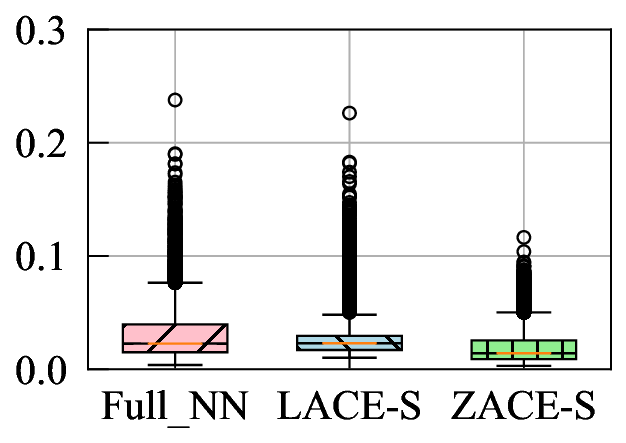}}
	\subfloat[Average deviation]{\includegraphics[scale=0.39]{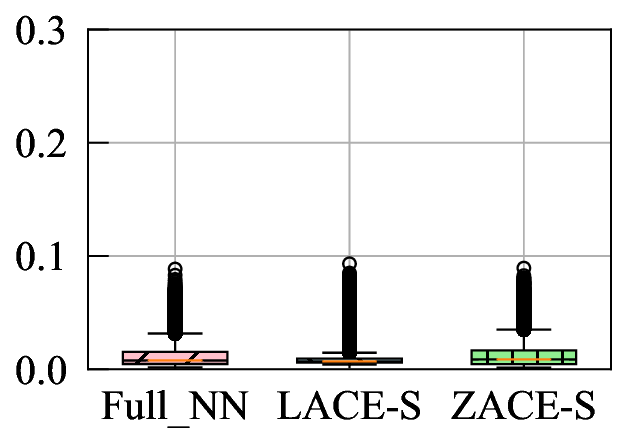}}
	\caption[]{\small Comparisons of the (a) maximum and (b) average absolute deviation in approximating the LMCE for the Full\_NN, LACE-S, and ZACE-S models using the 30-bus system.}\label{LACE_sim_2}
\end{figure}


We first demonstrate that our LACE-S model can successfully match the aggregated emission effects, achieving the total emission balance.
For this assessment, we use the error vector between the original NN output $\hat{\bblambda}$ and its projection $\tilde{\bblambda}$ [cf.~\eqref{projection}], namely the vector $|\hat{\bblambda}-\tilde{\bblambda}|$. This error vector quantifies the projection adjustment at each location and offers a more granular metric than the total balance loss in \eqref{eq:balanceloss}.
Fig.~\ref{LACE_sim_1} shows the box plots of the (a) maximum and (b) average entries of this distortion vector across load scenarios.
Each box is indicated by the first (Q1) and third (Q3) quartiles, along with the median midline. Data points beyond 150\% of the interquartile range (Q1--Q3) are shown as outliers.
Both models achieve nearly identical balance accuracy at an operationally negligible level, while the LACE-S tends to require fewer extreme corrections across scenarios.
In particular, LACE-S attains an average deviation of 0.005 and the maximum deviation of 0.008, comparable to the Full\_NN baseline. Given that the largest emission factor in Table \ref{tab:fuel_sim} is 0.9143, the distortion level below 1\% indicates an almost perfect emission balance.
This very high accuracy is attributed to \textbf{Stage 1} of the multi-stage training procedure in Section~\ref{sec:procedure}, which initializes the trainable parameters near the balanced baseline.
%

Beyond the balancing ability, the proposed LACE-S yields a clear improvement over the Full\_NN in approximating the marginal sensitivities, thanks to its ability to capture the distinct sensitivity patterns within each cluster.
Fig.~\ref{LACE_sim_2} shows the box plots for the absolute LMCE prediction error, namely $|\hat \bbmu - \bbmu|$, with both maximum and average statistics. 
Clearly, the improvement attained by our proposed LACE-S is two-fold. First, it can significantly reduce the maximum deviation from the Full\_NN's 0.08 (8\%) to below 0.04 (4\%). Second, the approximation performance across different load scenarios becomes more consistent, with the interquartile range reduced to roughly half of the baseline model.

\begin{figure}[t!]
	\centering
	\subfloat[Full\_NN]{\includegraphics[scale=0.26]{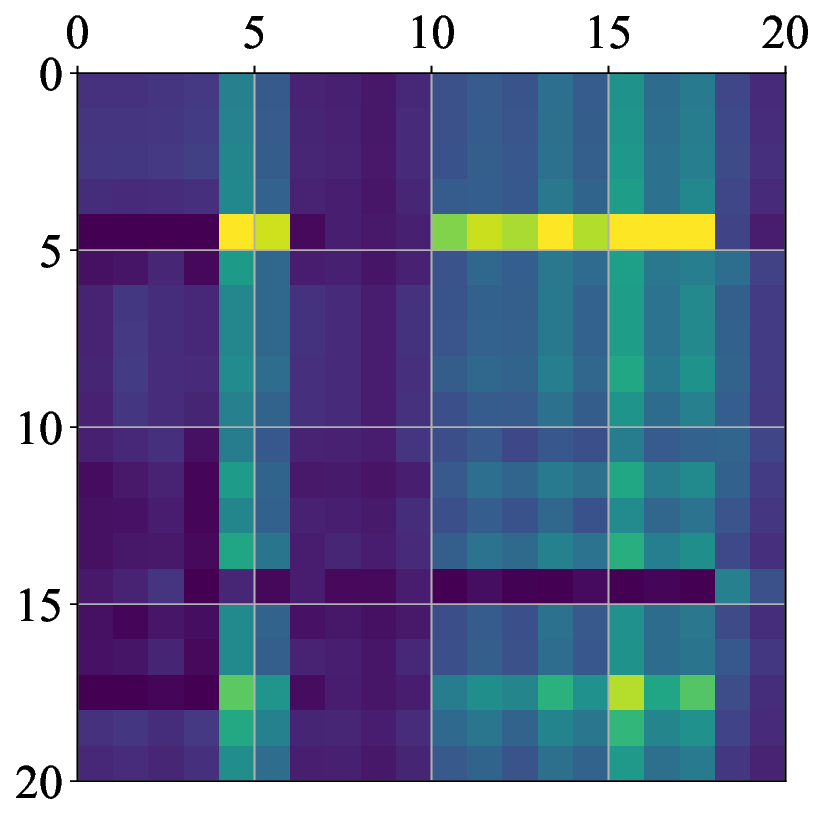}}\quad
    \subfloat[LACE-S]{\includegraphics[scale=0.26]{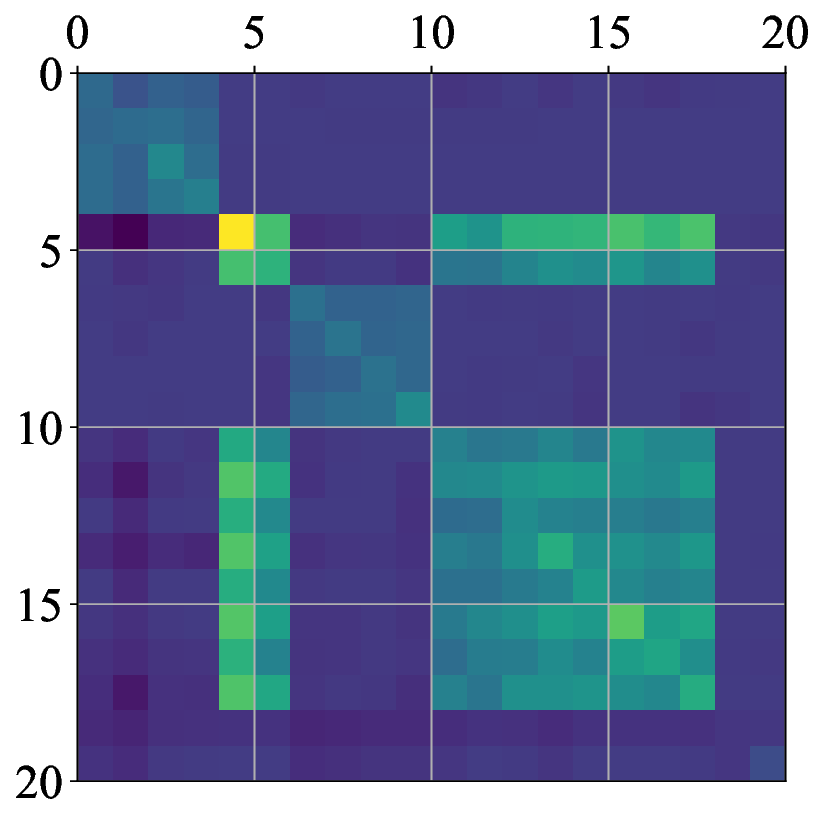}}
    \subfloat{\includegraphics[scale=0.3]{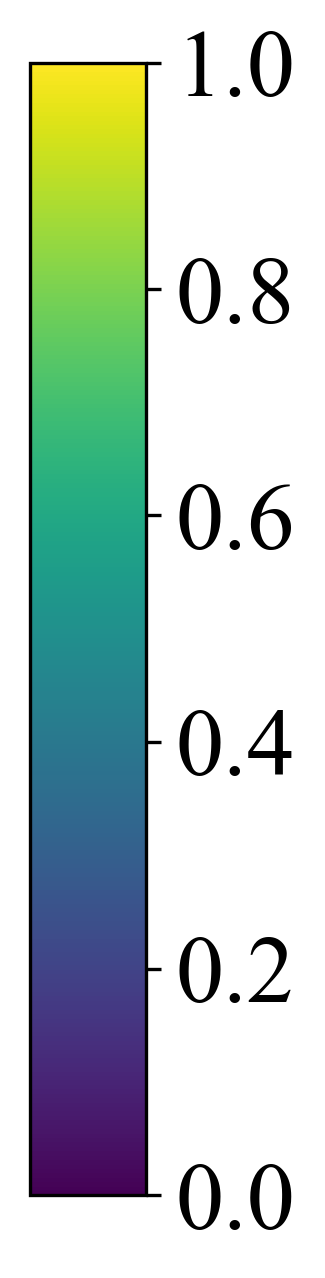}}
	\caption[]{\small Comparisons of Jacobian matrix of the (a) Full\_NN and (b) LACE-S models using a load profile of 120\% of nominal values on the 30-bus system.}\label{LACE_Jacobian}
\end{figure}

The superiority of the proposed LACE-S is further evidenced by the structure of the learned Jacobian matrix $\bbJ$. Fig.~\ref{LACE_Jacobian} compares the heatmaps of $\bbJ$ under a representative scenario with all loads at 120\% of their nominal values.
Without using the regularization step in Section~\ref{sec:LACEreg}, the baseline Full\_NN shows its Jacobian to have several column-wise blocks (e.g., loads 5--6 and 10--18). This observation implies that perturbations at these loads would have a global impact on emission factors throughout the system. Such a relation violates the partitioned structure expected from market re-dispatch, where any load perturbation practically affects a limited set of buses sharing the same marginal generators.
In contrast, our proposed LACE-S has successfully eliminated this global impact and produced four distinct blocks consistent with the block-diagonal structure in \eqref{loss_gamma1}. Since the four clusters have been obtained on the basis of LMCE similarity, the resultant Jacobian structure correctly limits the emission impact within the cluster of loads sharing the same marginal generators.
Moreover, diagonal entries have been successfully promoted by $\mathcal{L}_{\mathrm d}$ in \eqref{loss_gamma2}, ensuring that any load perturbation dominates its own location. Thus, these observations validate that our LACE-S aligns well with the intra-cluster coupling and self-influence structures for consistent carbon allocation.


\subsection{ZACE-S Modeling Performance}

We evaluate the ZACE-S model to assess whether the zonal aggregation in Section~\ref{sec:ZACE} can improve scalability without degrading model fidelity. We partition the 20 loads of the IEEE 30-bus system into $K=5$ predefined zones, following the zonal pricing structure based on LMP similarity~\cite{gebrekiros2015assessment,gebrekiros2015balancing}.
Accordingly, ZACE-S outputs the zonal emission factor $\bblambda^{\mathrm{z}} \in \mathbb R^5$, and we reduce the number of hidden neurons per layer to 30 to reflect the lower-dimensional output.

Figs.~\ref{LACE_sim_1} and~\ref{LACE_sim_2} have included the corresponding ZACE-S results under the same balance and sensitivity metrics. The NN model for ZACE-S has achieved comparable performance to the LACE-S one in terms of both emission balance and marginal sensitivity. Note that the sensitivity match for ZACE-S is based on the aggregated ZMCE in \eqref{eq:zmce}, not the original LMCE.
But still, the high accuracy achieved by ZACE-S has verified that its lower-dimensional carbon representation retains the dominant spatial patterns at the level of market zones. In particular, having a smaller number of neurons and output labels helps ZACE-S to reduce the NN model complexity. As summarized in Table~\ref{tab:computation}, the number of trainable parameters decreases from 3,200 for LACE-S to 1,650 for ZACE-S. This 51.56\% decrease effectively reduces the computational cost of NN training and Jacobian evaluation. Hence,  the proposed ZACE-S offers a scalable alternative for large power systems while preserving the high modeling accuracy at the zonal level.

\begin{table}[t!]
\centering
\caption{Training statistics for LACE-S and ZACE-S}
\label{tab:computation}
\begin{tabular}{l|cc}
\hline
Metric                                & LACE-S       & ZACE-S  \\ 
\hline
Trainable parameters                  & 3,200        & 1,650  \\ 
Average training time per epoch (s)   & 75.8  s      & 27.1 s  \\ 
Maximum training time per epoch (s)   & 151.3  s      & 74.8 s  \\ 
\hline
\end{tabular}
\end{table}


\subsection{Spatial Load Shifting Performance} \label{sec:sls}


The most important practical value of locational emission factors lies in determining carbon-aware demand-side adjustments. Specifically, spatial load shifting (SLS)~\cite{lindberg2021guide,lindberg2022using,gorka2025electricityemissions} utilizes these metrics to guide the spatial re-distribution of flexible loads. By explicitly relocating demand from carbon-intensive nodes toward cleaner locations, this operation dispatches the generation mix to reduce total system emissions. We evaluate the proposed LACE-S metric for SLS to demonstrate its efficacy as a globally consistent signal for accurately representing the potential reduction of carbon emissions for possible demand shifts.

%

To rigorously quantify this performance, we benchmark LACE-S guided load shifts against both a theoretical maximum and existing metrics. The optimal load shifting (Opt-shift) problem computes the absolute upper bound on emission reduction by assuming perfect operator knowledge of load flexibility. Against this theoretical limit, we compare signal-based SLS solutions driven by LACE-S, LMCE, LACE-R, and carbon emission flow (CEF) metrics~\cite{kang2012carbon,kang2015carbon}. The detailed mathematical formulations for these problems are provided in the Appendix.
The optimization problems are implemented in Pyomo~\cite{hart2011pyomo} and solved with the Gurobi optimizer~\cite{gurobi}. Using the 30-bus test system, we select six high-loading buses ($\mathcal{S}=\{2,7,8,12,19,21\}$) as flexible loads and allow up to 5 MW of shiftable demand at each bus. 
To capture these perturbations, we expand the training dataset of LACE-S to encompass the range of shiftable load demands.
For the signal-based SLS, we first solve the SLS problem~\eqref{sls} with a given metric to obtain the decision vector of shifted load, and then re-run the DCOPF problem with this updated load profile to compute the \textit{realized} post-shift total emissions. The second step ensures a fair comparison among signal-based SLS solutions, because the optimal objective provided by \eqref{sls} alone is only an estimate, not the actual emission reduction. 

\begin{table}[t!]
\centering
\caption{Comparison of CO$_2$ emissions and reductions by carbon emission metrics}
\label{tab:co2_reductions}
\begin{tabular}{l|cc}
\hline
Metric        & Emissions (tCO$_2$) & Change (tCO$_2$) \\ 
\hline
Pre-shift       & 176.062            & ---                 \\ 
Opt-shift       & 175.829            & -0.233 (-0.13\%)              \\ 
\hline
\textbf{LACE-S}  & \textbf{175.887} & \textbf{-0.175 (-0.10\%)}   \\
LMCE           & 176.286             & +0.224 (+0.13\%)               \\ 
LACE-R         & 176.156             & +0.094 (+0.05\%)              \\ 
CEF            & 176.286             & +0.224 (+0.13\%)               \\
\hline
\end{tabular}
\end{table}

We first use a single loading scenario with all loads scaled to 120\% of their nominal values to compare the metrics as SLS signals.
Table~\ref{tab:co2_reductions} summarizes the pre-shift emissions, the Opt-shift benchmark, and the \textit{realized} post-shift emissions for signal-based SLS solutions using each metric. Note that the change in emissions and its percentage are in comparison with the pre-shift baseline. 
Our proposed LACE-S has achieved the highest reduction among all signal-based SLS solutions, as indicated by the \textbf{bold} font. 
In fact, LACE-S is the \textit{only} metric that has resulted in emission reduction for this loading profile: it lowers the total emissions from 176.062 to 175.887 tCO$_2$, corresponding to a 0.10\% reduction of the pre-shift emissions. By contrast, LMCE and LACE-R have respectively increased the emissions by 0.13\% and 0.05\%; and similarly for CEF with a 0.13\% increase. Hence, these three existing metrics have unfortunately led the SLS to a higher emission profile. Note that although Opt-shift has produced a higher reduction than our LACE-S, the former requires the knowledge of unavailable information and cannot be implemented in practice. Interestingly, the LACE-S based SLS is very close to this theoretical benchmark, demonstrating the superior SLS performance attained by our proposed metric.  

\begin{figure}[t!]
	\centering
	\subfloat[Opt-shift]{\includegraphics[scale=0.35]{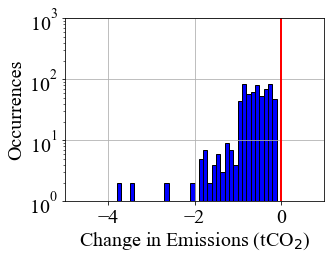}}\quad
    \subfloat[LACE-S]{\includegraphics[scale=0.35]{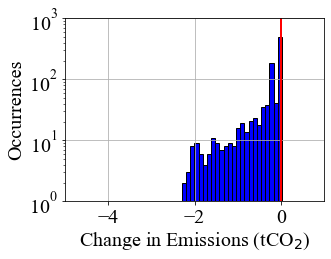}}\quad
    \subfloat[LMCE]{\includegraphics[scale=0.35]{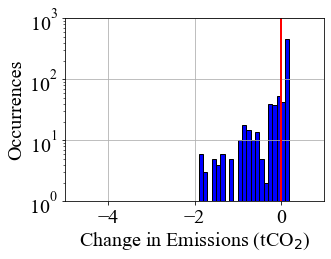}}\quad
    \subfloat[LACE-R]{\includegraphics[scale=0.35]{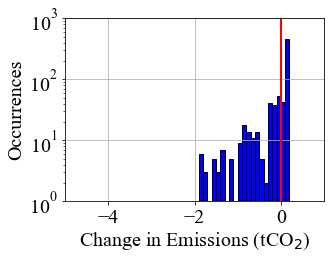}}\quad
    \subfloat[CEF]{\includegraphics[scale=0.35]{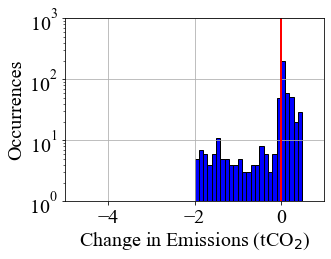}}
	\caption[]{\small Histograms of realized emission changes across 1,000 load profiles for (a) Opt-shift and (b)-(e) metric-based SLS.}\label{fig:SLS_result}
\end{figure}

To expand this comparison beyond a single profile, we plot the histograms of the emission changes for all these methods across 1,000 distinct load profiles in Fig.~\ref{fig:SLS_result}. In all plots, the red line marks the reference of no emission change. Thus, the further the samples are below the red line, the more emissions are reduced. As expected, the Opt-shift solution always yields negative emission changes, while the LACE-S based SLS attains performance very close to this theoretical benchmark and never increases the post-shift emissions. In some samples, the latter results in zero emission changes, but it never increases the total emissions beyond the red line.
On the contrary, the three other metrics, LMCE, LACE-R, and CEF, all have a significant number of samples beyond the red line, and skew the distributions toward the positive range. LMCE and LACE-R provide very similar distributions, with CEF-based SLS further increasing the occurrence of positive changes. This observation indicates that for many scenarios, using the three existing metrics has led to more carbon-intensive electricity production after the load shifting.
Together, the results in Table~\ref{tab:co2_reductions} and Fig.~\ref{fig:SLS_result} have systematically confirmed that our proposed LACE-S provides a reliable, sensitivity-consistent signal for carbon-aware SLS.

\section{Conclusions} \label{sec:con}
We developed a sensitivity-consistent metric for locational average carbon emissions (LACE-S) using a neural representation that maps the entire operating region. By providing a globally consistent mapping, this approach explicitly overcomes existing metrics that remain valid only around limited operating regions.
The neural modeling seeks to attain the total emission balance through an explicit projection layer, while a gradient-based loss matches the marginal emission sensitivities. Furthermore, Jacobian-based regularization is introduced to capture the underlying partition of load buses with closely aligned generator responses. We also defined the zonal counterpart, ZACE-S, to reduce the model complexity by using predefined market zones.
Numerical tests on the IEEE 30-bus system show the accurate match with emission balance and sensitivity attained by our neural LACE-S model. Crucially, our proposed LACE-S provides an effective signal for carbon-aware spatial load shifting and yields consistent emission reductions across the loading region, while existing metrics increase post-shift emissions for most scenarios.

Exciting future directions open up for our LACE-S metric based on neural representation, such as extensions to more realistic market models and nonlinear constraints.  Robustness under renewable uncertainty and temporal coupling can be investigated, especially for large-scale systems. Moreover, we will enhance the practical implementation of constructing the LACE-S metric, using limited samples or online learning to adapt to evolving conditions, to facilitate its integration into real-world market-design studies for carbon-aware operations.

\section*{Appendix}
\subsection{Optimal Load Shifting}\label{app:ols}
We begin with an \textit{optimal} load shifting (Opt-shift) benchmark, which characterizes the maximum emission reduction. It is based on the idealized assumption that the market operator has full knowledge of load flexibility and can directly dispatch flexible loads~\cite{lindberg2022using}. This yields a bi-level formulation with the upper-level identifying the optimal choice of load shifts that minimizes the resultant total emissions, while the lower-level computes the corresponding market dispatch via DC-OPF to satisfy all network constraints. For the set of flexible loads denoted by $\mathcal{S}$, the Opt-shift problem determines the post-shift demand $s_i$ from given nominal demand $d_i$ for each load $i \in \mathcal{S}$, as given by
\begin{subequations}\label{ols}
\begin{align}
    \min_{\bbs} ~ & \bbf^\top \bbg\\
    \textrm{s.t.} ~
    & (1-\Delta) d_i \leq s_i \leq (1+\Delta) d_i, ~\forall i \in \mathcal{S}\label{shift_range}\\
    & \textstyle \sum_{i\in\mathcal{S}} d_i = \textstyle \sum_{i\in\mathcal{S}} s_i\label{total_balance}\\
    & \min_{\underline\bbg \leq \bbg \leq \bar\bbg} ~ \bbc^\top \bbg \\
    &~\quad \textrm{s.t.} ~
    \bbA \bbg= \bbb(\bbd, \bbs),\label{ols_eq}\\
    &\qquad\quad \bbl(\bbd) \leq \bbB \bbg \leq \bbu(\bbd).\label{ols_ineq}
 \end{align}
\end{subequations}
In the upper level, the constraints in \eqref{shift_range} bound the shifting flexibility such that each $s_i$ is within a given fraction $\Delta$ of its nominal value $d_i$, while the equality in \eqref{total_balance} helps to maintain the same total demand to eliminate the possibility of simply reducing the total demand. The lower level is precisely the market-dispatch problem \eqref{opf} by accounting for the load shifting associated with the upper-level decisions $\bbs$. 
It is worth emphasizing that the Opt-shift problem \eqref{ols} relies on the non-realistic setting that market operators have the full information of load flexibility, and thus it provides an \textit{unachievable} upper bound on the total emission reductions that we use to benchmark the optimality gap attained by realistic SLS solutions later.

\subsection{Signal-based Spatial Load Shifting}\label{app:sls}
Real-world electricity markets adopt a carbon signal-based SLS formulation; see e.g. the overview in~\cite{gorka2025electricityemissions}. By determining the emission-associated factors denoted by vector $\bbphi$, the market operator can indirectly encourage individual loads to explore their respective flexibility for emission reduction.
Hence, all aforementioned metrics, such as LMCE, LACE-R, and our proposed LACE-S, can serve as a market signal after the initial clearing process.
Based on $\bbphi$, individual loads seek to minimize the estimated emissions, given by
\begin{subequations}\label{sls}
\begin{align}
  \hhatbbs=  \arg\min_{\bbs} ~ & \textstyle 
  \bbphi^\top  \bbs \label{sls-obj}\\
    \textrm{s.t.} ~
    & \eqref{shift_range} - \eqref{total_balance}.
 \end{align}
\end{subequations}
Since $\bbphi$ is computed from the pre-shift market-clearing solutions, the objective in \eqref{sls-obj} merely serves as a proxy for the true reduction. This way, the signal from the market operator and the response by the loads are decoupled to reflect market practices. To evaluate the realized impact of the SLS decisions by solving \eqref{sls}, we need to use $\hhatbbs$ and re-solve DC-OPF under this shifted load profile. This post-SLS procedure allows us to effectively compare the realizable reduction of emissions over SLS decisions under different choices of signal $\bbphi$.

\ifCLASSOPTIONcaptionsoff
\newpage
\fi
	
\bibliographystyle{IEEEtran}
\bibliography{Ref.bib}

\end{document}